\documentclass[prb,twoside,twocolumn,10pt,floatfix,showpacs,citeautoscript,letter]{revtex4-2}

\usepackage{amsmath}
\usepackage{amssymb}
\usepackage{glossaries}
\usepackage{graphicx}
\usepackage[version=4]{mhchem}
\usepackage{siunitx}
\usepackage{tabularx}
\usepackage{hyperref}

\newcommand \Title{
    Revealing the free energy landscape of halide perovskites: \texorpdfstring{\\}{}
    Metastability and transition characters in \texorpdfstring{\ce{CsPbBr3}}{CsPbBr3} and \texorpdfstring{\ce{MAPbI3}}{MAPbI3}
}

\hypersetup{
    pdfauthor={E. Fransson, J. M. Rahm, J. Wiktor, and P. Erhart},
    pdftitle={\Title},
    pdffitwindow=false,
    pdfstartview={FitH},
    pdfnewwindow=true,
    colorlinks=true,
    linkcolor=black,
    citecolor=black,
    filecolor=black,
    urlcolor=black,
}

\setacronymstyle{long-short}
\newacronym{acf}{ACF}{autocorrelation function}
\newacronym{dft}{DFT}{density functional theory}
\newacronym{rms}{RMS}{root mean square}
\newacronym{gpu}{GPU}{graphics processing unit}
\newacronym{md}{MD}{molecular dynamics}
\newacronym{mlp}{MLP}{machine-learned potential}
\newacronym{nep}{NEP}{neuroevolution potential}
\newacronym{si}{SI}{Supporting Information}
\newacronym{soap}{SOAP}{smooth overlap of atomic positions}
\newacronym{xc}{XC}{exchange-correlation}
\newacronym{scan}{SCAN}{strongly constrained and appropriately normed}
\newacronym{wham}{WHAM}{weighted histogram analysis method}

\renewcommand{\vec}[1]{\boldsymbol{#1}}

\DeclareSIUnit\angstrom{\text{Å}}
\DeclareSIUnit\atom{\text{atom}}

\global\let\oldnewlabel\newlabel
\gdef\newlabel#1#2{\newlabelxx{#1}#2}
\gdef\newlabelxx#1#2#3#4#5#6{\oldnewlabel{#1}{{#2}{#3}}}
\let\newlabel\oldnewlabel
\newcommand{\addchalmers}{Department of Physics, Chalmers University of Technology, SE-41296, Gothenburg, Sweden}

\begin{document}

\title{\Title}

\author{Erik Fransson}
\altaffiliation{These authors contributed equally to this work}
\author{J. Magnus Rahm}
\altaffiliation{These authors contributed equally to this work}
\author{Julia Wiktor}
\author{Paul Erhart}
\affiliation{\addchalmers}
\email{erhart@chalmers.se}

\begin{abstract}
Halide perovskites have emerged as a promising class of materials for photovoltaic applications.
A challenge in these applications is how to prevent the crystal structure from degradation to photovoltaically inactive phases, which requires an understanding of the free energy landscape of these materials.
Here, we uncover the free energy landscape of two prototypical halide perovskites, \ce{CsPbBr3} and \ce{MAPbI3} via atomic scale simulations using umbrella sampling and machine-learned potentials.
For \ce{CsPbBr3} we find very small free energy differences and barriers close to the transition temperatures for both the tetragonal-to-cubic and the orthorhombic-to-tetragonal transition.
For \ce{MAPbI3}, however, the situation is more intricate.
In particular the orthorhombic-to-tetragonal transition exhibits a large free energy barrier and there are several competing tetragonal phases.
Using large-scale molecular dynamics simulations we explore the character of these transition and observe latent heat and a discrete change in structural parameters for the tetragonal-to-cubic phase transition in both \ce{CsPbBr3} and \ce{MAPbI3} indicating first-order transitions.
We find that in \ce{MAPbI3} the orthorhombic phase has an extended metastability range and furthermore identify a second metastable tetragonal phase.
Finally, we compile a phase diagram for \ce{MAPbI3} that includes potential metastable phases.
\end{abstract}

\maketitle

\section{Introduction}

Halide perovskites are highly promising materials for photovoltaic applications, with demonstrated efficiencies exceeding 25\% \cite{NREL}.
Their application potential is, however, still limited by stability issues, which has prompted extensive research into the dynamics of their crystal structure~\cite{frost2016moving, carignano2017critical, jinnouchi2019phase,kaiser2021first, baldwin2023dynamic, WikFraKub23} and the stability of different perovskite phases \cite{yi2016entropic, beal2016cesium, jinnouchi2019phase, raval2022understanding}.

In spite of the very large body of literature available, halide perovskites continue to surprise.
For example, a previously not experimentally recognized crystal structure of the the widely studied perovskite \ce{CsPbBr3}, was detected in a combination of convergent beam electron diffraction and electron ptychography \cite{dos2018determination}.
Additionally, several recent studies showed that short-range ordering in octahedral tilt angles characteristic of low-temperature phases persists even in the cubic phase of halide perovskites \cite{weadock2023nature, WikFraKub23, baldwin2023dynamic}.
Furthermore, the widely accepted low-temperature Pnma structure of \ce{MAPbI3} has been put into question by a combination of inelastic neutron scattering and first-principles calculations \cite{druzbicki2016unexpected}. 
As an alternative the authors proposed that structures with lower local symmetry are needed to reproduce experimental data \cite{druzbicki2016unexpected} and in fact several halide perovskite structures have been shown to be locally rather similar \cite{cannelli2022atomic}.
The structural models differ, however, in properties that are crucial for photovoltaic applications, most importantly the band gap \cite{wang2017indirect}.
Therefore, a more detailed understanding of the local atomic structure and dynamics as well as their coupling to phase transitions is needed for further optimizing these materials.

Similarly, the character of the phase transitions in halide perovskite has been debated.
Both the tetragonal-to-cubic and orthorhombic-to-tetragonal phase transitions are driven by soft phonon modes, specifically shear modes at the M and R-points on the Brillouin zone boundary of the cubic structure, which in the following we simply refer to as M and R modes for brevity.
For a purely continuous transition the frequency of the soft modes should go smoothly to zero at the transition temperature.
In \ce{CsPbBr3} the tetragonal-to-cubic phase transition has, however, been observed to have first-order character through experimental studies \cite{Hirotsu1974, Rodov2003, Stoumpos2013}, whereas the transition between the tetragonal and orthorhombic phases has been characterized as purely continuous \cite{Hirotsu1974}.
In \ce{MAPbI3} the character of the tetragonal-to-cubic transition has been analyzed in numerous experimental studies with different conclusions concerning the order of the transition \cite{OnodaYamamuro1990, Kawamura2002, Stoumpos2013b, Weller2015, Whitfield2016}, although most studies point towards it being first order.
The orthorhombic-to-tetragonal transition in \ce{MAPbI3} has been consistently observed to have first-order character.

One should also note that mixing both on the cation and anion sites involving two or more different species has been demonstrated as a promising route for improving the thermodynamic stability of these materials \cite{mcmeekin2016mixed, xu2017mixed, kubicki2018formation, fan2019review}.
This renders the configuration (and phase) space of these materials even more complex as it adds further compositional degrees of freedom.

From a computational standpoint, the very dynamic crystal structure of halide perovskites \cite{egger2016hybrid, Monacelli2023} presents a challenge.
Specifically, vibrational properties play a key role in determining the free energy landscape and ground state \gls{dft} calculations provide only very limited information in this regard.
To capture the dynamic properties, ab-initio \gls{md} simulations are a popular approach, and have shone light on, among other things, the band gaps of halide perovskites \cite{carignano2015thermal, wiktor2017predictive, mladenovic2018effects, zhu2022probing, gebhardt2022electronic}, local disorder \cite{girdzis2020revealing, cannelli2022atomic}, and defect energy levels \cite{cohen2019breakdown}.
Yet, the computational cost of ab-initio \gls{md} simulations prevents extensive sampling and thereby limits access to structural and dynamic correlations that require careful convergence.
Since \glspl{mlp} have the ability to significantly speed up sampling with a negligible loss in accuracy, there has been a number of applications of these techniques to halide perovskites \cite{jinnouchi2019phase, lahnsteiner2019long, thomas2019machine, zhou2020structural, mangan2021dependence, bokdam2021exploring, gruninger2021microscopic, lahnsteiner2022anharmonic, BraGoeVan22, FraRosEri23, FraWikErh2023, WikFraKub23}.
For example, Jinnouchi \textit{et al.} \cite{jinnouchi2019phase} constructed a \gls{mlp} for \ce{MAPbI3} to study the phase transitions \cite{jinnouchi2019phase}.
However, the previous studies focused on specific, predetermined phases or transition between those.
Therefore, our understanding of the free energy landscape of halide perovskites is still evolving.

Here, we uncover the free energy landscape of two halide perovskites, \ce{CsPbBr3} and \ce{MAPbI3} with respect to the tilting degrees of freedom, by constructing \gls{mlp} that are both accurate and computationally very efficient on \glspl{gpu} \cite{FanWanYin22, FraWikErh2023}.
The structural phase space is explored via \gls{md} simulations and  umbrella sampling \cite{FanWeiVie2017}.
We demonstrate that the free energy minima in \ce{CsPbBr3} for the various phases are typically wide and soft, with small free energy differences between phases.
For \ce{MAPbI3} the free energy landscape indicates a large barrier between the orthorhombic and tetragonal phases as well as several competing tetragonal phases.
We relate these results to the available tilt modes by analyzing the space of accessible tilt patterns using the concept of ``Glazer space''.
This provides a geometrical view of the possible transitions that is transferable beyond the specific compounds and chemistries considered here.
Our analysis enables us to clearly resolve the character of the different phase transitions in these systems in agreement with the majority of experimental assignments and eventually to construct a phase diagram for \ce{MAPbI3} that includes potential metastable phases.

\section{Methods}

\subsection{Density functional theory calculations}
\Gls{dft} calculations were performed using the projector augmented-wave method  \cite{Blo94} as implemented in the Vienna ab-initio simulation package \cite{KreHaf93, KreFur96}.
The exchange-correlation contribution was represented using the \gls{scan} density functional \cite{SunRuzPer15}, which has been previously established for the study of these systems \cite{Lahnsteiner2016dynamics, jinnouchi2019phase}.
The Brillouin zone was sampled using a $\Gamma$-centered grid with a $\vec{k}$-point density of \SI{0.25}{\per\angstrom} and Gaussian smearing with a width of \SI{0.1}{\electronvolt}.

\subsection{Neuroevolution potentials}
For \ce{CsPbBr3} we employed the \gls{nep} model described in our recent work \cite{FraRosEri23}, and for \ce{MAPbI3} we constructed a model by similar means using the \textsc{gpumd} package \cite{FanWeiVie2017, FanZenZha21, FanWanYin22} in conjunction with the \textsc{calorine} \cite{calorine} and \textsc{ase} packages \cite{Larsen2017}.
The models are based on a neural network for which local atomic environments are described by radial and angular components \cite{FanZenZha21}.
In the NEP3 version \cite{FanWanYin22} employed here the radial part of the atomic environment descriptor is constructed from linear combinations of Chebyshev basis functions while the three-body angular part is similarly build from Legendre polynomials.

Here, we used radial and angular cutoffs of \SI{8}{\angstrom} and \SI{4}{\angstrom}, respectively, and terminated the radial and angular expansions at orders 12 and 6, respectively.
The neural network consisted of one hidden layer of 40 neurons with a hyperbolic tangent activation function and was trained using the natural evolution strategy \cite{WieSchGla14} implemented in \textsc{gpumd}, using \num{200000} and \num{500000} generations for training the \ce{CsPbBr3} and \ce{MAPbI3} models, respectively.
Both $\ell_1$ and the $\ell_2$ regularization terms were included, using $\lambda_1=\lambda_2=0.02$ for \ce{CsPbBr3} and $\lambda_1=\lambda_2=0.05$ for \ce{MAPbI3} (with $\lambda_1$ and $\lambda_2$ as defined in Ref.~\citenum{FanWanYin22}).
Structures for training and validation were generated by means of a bootstrapping approach, i.e., first preliminary models were fitted, new structures were extracted from \gls{md} simulations based on these models, and then used to fit the next model generation, similar to the approach outlined in Ref.~\citenum{FraWikErh2023}.
The final model for \ce{CsPbBr3} was fitted to forces, energies, and virials from \gls{dft} calculations for \num{642} atomic structures with a total of \num{176920} atoms, and validated against a set of \num{72} structures with a total of \num{19800} atoms.
In this holdout set, the model achieved an \gls{rms} error of \SI{0.6}{\milli\electronvolt\per\atom} for energies, \SI{15}{\milli\electronvolt\per\atom} for virials, and \SI{43}{\milli\electronvolt\per\angstrom} for forces.
The \ce{MAPbI3} model was fitted to forces, energies, and virials from \gls{dft} calculation of \num{511} atomic structures with a total of \num{172092} atoms, and validated against a set of \num{57} structures with a total of \num{18804} atoms.
It achieved a \gls{rms} error of \SI{0.6}{\milli\electronvolt\per\atom} for energies, \SI{4.5}{\milli\electronvolt\per\atom} for virials, and \SI{38}{\milli\electronvolt\per\angstrom} for forces (see \gls{si} for details).

\begin{figure*}
    \includegraphics[scale=0.825]{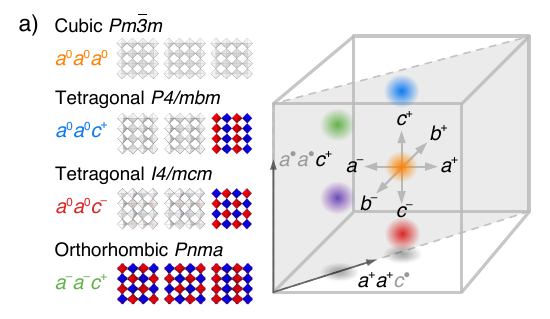}
    \includegraphics{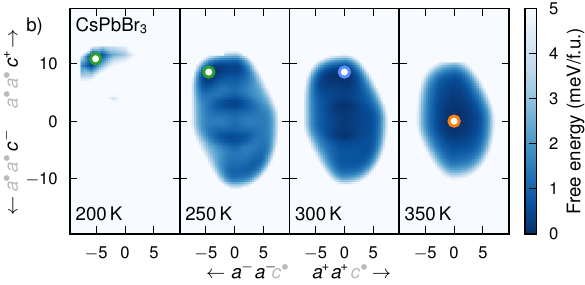}
    \vspace{-14pt}
    \begin{flushright}
    \includegraphics{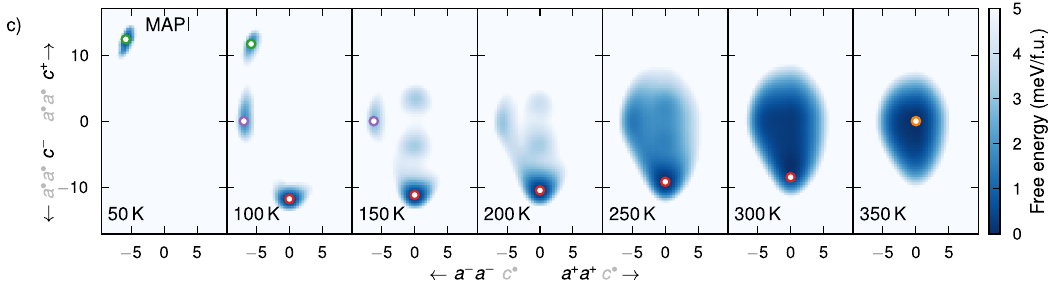}
    \end{flushright}
    \caption{
    (a) Commonly occurring perovskite phases and their positions in Glazer space.
    The insets on the left-hand side show the tilting patterns along the $x$, $y$, and $z$ directions, where red and blue indicate positive and negative tilt angles, respectively, whereas gray represents the absence of tilting.
    In addition to the four experimentally observed phases shown on the left, the figure also indicates the position of the $a^-a^-c^0$ structure (in purple), which can be observed on heating in \ce{MAPbI3} if the first-order transition from $a^-a^-c^+$ to $a^0a^0c^-$ is kinetically suppressed (see \autoref{fig:CsPbBr3-MAPI-heating-cooling}c).
    The atomic structures were visualized using \textsc{ovito} \cite{Stukowski2010}.
    (b,c) Free energy landscape of (b) \ce{CsPbBr3} and (c) \ce{MAPbI3} in Glazer space $a^\bullet b^\bullet c^\bullet$ with $a=b$.
    The abscissa in the free energy heat maps corresponds to the tilt angle in degrees about both $x$ and $y$-axes, where negative and positive values indicate out-of-phase and in-phase tilting, respectively.
    The ordinate is the tilt angle about the $z$-axis.
    }
    \label{fig:glazer-space-free-energy}
\end{figure*}

\subsection{Molecular dynamics and umbrella sampling}
\Gls{md} simulations were carried out using the \textsc{gpumd} package (version 3.2) \cite{FanWeiVie2017} using a custom patch for umbrella sampling.
In these simulations, time was propagated in steps of \SI{0.5}{\femto\second} in systems containing hydrogen, otherwise a time step of \SI{1}{\femto\second} was employed.
The Berendsen thermostat and barostat \cite{BerPosGun84} were used to control temperature and pressure.
Systems were initialized from \numproduct{4x4x3} supercells of the ideal primitive orthorhombic structure, for a total of \num{192} formula units.
The system size was chosen to be large enough to avoid finite-size effects but small enough to exhibit the statistical fluctuations required for free energy integration with \gls{wham} (see below). 
Furthermore, the global constraints in the free energy sampling become weaker for larger system sizes and the system may not stay in the desired crystal structure but instead form interfaces or domains.
The cell was constrained to remain orthorhombic, i.e., the volume was free to fluctuate but the angles between cell vector were fixed to \SI{90}{\degree}, throughout the simulations to simplify the analysis.

Umbrella sampling \cite{Torrie1977} was used to extract the free energy in ``Glazer space'' (\autoref{fig:glazer-space-free-energy}a and \autoref{sect:glazer-space}).
To this end, we biased the simulations using six collective variables: projections $\xi$ of atomic displacements on the three M modes and the three R modes as done in Ref.~\citenum{FraRosEri23}.
Each phonon mode projection was biased with a quadratic energy term,
\begin{equation*}
    U_\text{mode} = k (\xi_\text{mode} - \hat{\xi}_\text{mode})^2,
\end{equation*}
where $\xi_\text{mode}$ is the projection on the respective phonon mode displacement (with the full displacement vector normalized to 1), $\hat{\xi}_\text{mode}$ is a specific target projection, and the spring constant $k=\SI{0.1}{\per\electronvolt}$.
We performed one \gls{md} simulation of \SI{1}{\nano\second} for each combination of values of $\hat{\xi}_\text{mode}$ in order to span the Glazer space along the diagonal $x=y$.
If the target projection of any of the modes M$_i$ was non-zero, the corresponding R$_i$ target projection was set to zero.
Free energies were calculated with \gls{wham} \cite{KumRosBou95} using the software provided by Grossfield \cite{wham-Grossfield} but extended to handle the three-dimensional case.
At low temperatures ($\leq \SI{200}{\kelvin}$ for \ce{CsPbBr3}, $\leq \SI{50}{\kelvin}$ for \ce{MAPbI3}) we resorted to umbrella integration since fluctuations were too small to achieve sufficiently overlapping histograms from individual simulations.

Mapping the free energy across the Glazer space thus allows one to identify the phases corresponding to the free energy minima, their respective  tilt angles, and gain insight regarding the barriers between phases \cite{Klarbring2019}.

\section{Results and discussion}

\subsection{The Glazer space}
\label{sect:glazer-space}

The perovskite crystal structure has the general formula \ce{ABX3}.
The B atoms reside in corner-sharing octahedra formed by the X atoms, which in the case of halide perovskites are halide species.
These octahedra tilt according to patterns that are distinctive for each phase.
Typically, there are at least three phases that are stable in different temperature intervals: an orthorhombic ground state, a tetragonal phase, and a cubic phase.
The tilt patterns are associated with phonon instabilities.
In this regard the three M modes and the three R modes of the cubic structure are particularly important \cite{FraRosEri23}.
Each M mode corresponds to the rotation of an octahedron around one of the three major axes, with consecutive octahedra along that axis rotated in-phase.
The R modes have a similar effect but the octahedra rotate out-of-phase.
The different perovskite crystal structures are conveniently described by the projections on these modes, for which Glazer introduced a compact notation \cite{Glazer1972}.
For example, the commonly occurring Pnma phase can be written as $a^- a^- c^+$, which corresponds to a structure with an out-of-phase rotation of $a$ degrees about both the $x$ and $y$ axes (the R$_x$ and R$_y$ modes) and an in-phase rotation of $c$ degrees about the $z$ axis (M$_z$).
The cubic phase, on the other hand, corresponds to $a^0 a^0 a^0$ (no average rotation about any axis).
We will refer to the three-dimensional space spanned by in-phase and out-of-phase rotation about the three axes as the \emph{Glazer space} (see \autoref{fig:glazer-space-free-energy}a for a schematic).
All known, stable perovskite phases of \ce{CsPbBr3} and \ce{MAPbI3} correspond to distinct regions in Glazer space.

\begin{figure*}
    \centering
    \includegraphics{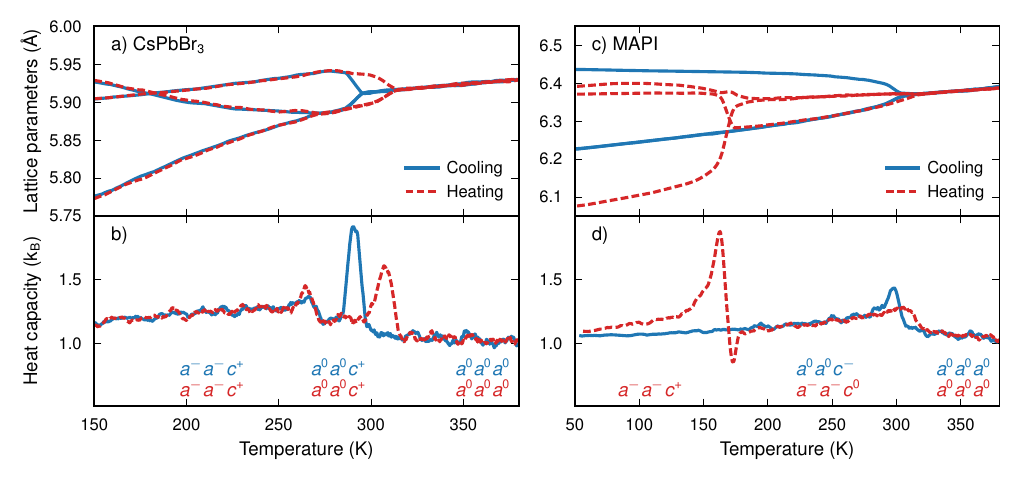}
    \caption{
    Lattice parameters and heat capacity from heating and cooling \gls{md} simulations in the $NPT$ ensemble for (a,b) \ce{CsPbBr3} and (c,d) \ce{MAPbI3}.
    The heat capacity is reported per degree of freedom in the system.
    }
    \label{fig:CsPbBr3-MAPI-heating-cooling}
\end{figure*}

\subsection{\texorpdfstring{\ce{CsPbBr3}}{CsPbBr3}}

\textbf{Behavior on heating/cooling.}
First, we run heating and cooling \gls{md} simulations in the $NPT$ ensemble for \ce{CsPbBr3} with \num{61440} atoms and a cooling/heating rate of about \SI{10}{\kelvin\per\nano\second} (\autoref{fig:CsPbBr3-MAPI-heating-cooling}a,b).
Two phase transitions are observed, at around \SI{260}{\kelvin} and \SI{300}{\kelvin}, corresponding to the orthorhombic-to-tetragonal transition and the tetragonal-to-cubic transition, respectively.
The transitions are clearly visible as changes in the lattice parameters, but can also be determined from peaks and kinks in the heat capacity \cite{FraWikErh2023}.
For the tetragonal-to-cubic transition some hysteresis is observed suggesting first-order character, whereas for the orthorhombic-to-tetragonal transition no hysteresis is observed, fully in line with a continuous phase transition.

\textbf{Free energy landscape in Glazer space.}
It is now instructive to consider the temperature dependence of the free energy landscape in Glazer space (\autoref{fig:glazer-space-free-energy}).
At \SI{350}{\kelvin} the free energy minimum is found at the origin of the Glazer space, i.e., the cubic $a^0a^0a^0$ structure (Pm$\bar{3}$m).
At \SI{300}{\kelvin} the free energy is lowest for the tetragonal $a^0a^0c^+$ structure (P4/mbm), although the free energy landscape is very flat along the $z$-tilt axis towards the cubic phase.
The tilt angle of the tetragonal structure is about \SI{8}{\degree}.
At \SI{250}{\kelvin} the free energy minimum is found for the orthorhombic $a^-a^-c^+$ structure (Pnma).
Finally, at \SI{200}{\kelvin} a distinct narrow free energy minimum is obtained for the $a^-a^-c^+$ phase, which continues to be the stable phase down to \SI{0}{\kelvin}.
The tilt angle about the $z$-axis is the same as in the tetragonal structure, while the tilt angle about $x$ and $y$ is about \SI{4}{\degree}.
Again, we note that the free energy landscape between the orthorhombic ($a^-a^-c^+$) and the tetragonal ($a^0a^0c^+$) phase is very flat.
The local free energy minimum closer to the center of Glazer space corresponds to the $a^0 a^0 c^+$ phase but with multiple domains separated by anti-phase boundaries, which arises due to the global constraint on the phonon mode coordinate.
These can for example occur in a \numproduct{6x6x6} (cubic) supercell with four layers of primitive cells tilting with the same phase, while the other two tilt with the opposite phase.
The resulting structure has a global order parameter (mode coordinate) equal to $1/3$ of the $a^0 a^0 c^+$ ground-state.

The calculated free energy in Glazer space contains the correct phase transitions as observed in experiments \cite{Hirotsu1974, Sharma1991, Rodov2003, Stoumpos2013, Lopez2020, Svirskas2020, Malyshkin2020}.
They occur, however, at slightly lower temperatures, which can be attributed to the exchange-correlation functional used in the reference \gls{dft} calculations \cite{FraWikErh2023}.

\begin{figure*}
    \centering
    \includegraphics{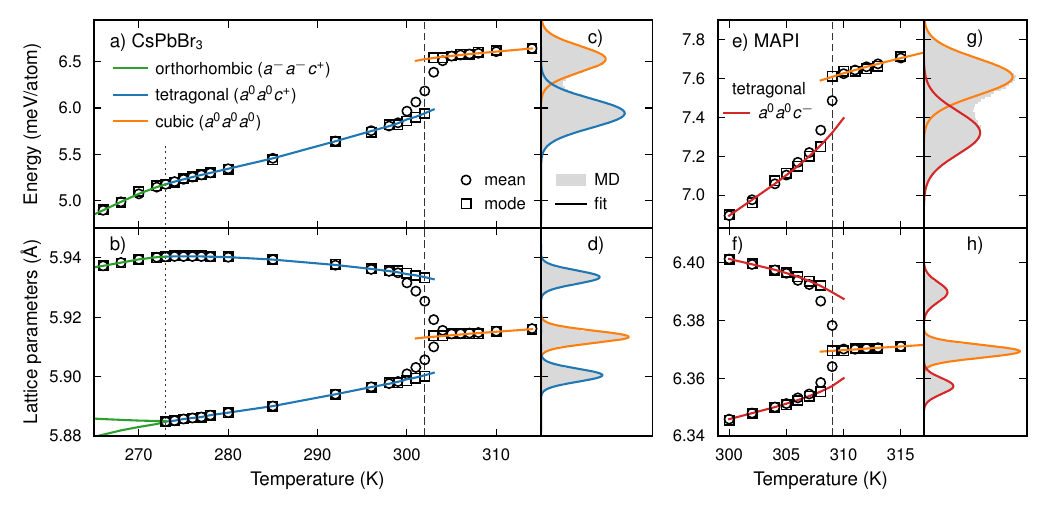}
    \caption{
    (a) Energy and (b) lattice parameters from \gls{md} simulations in the $NPT$ ensemble around the cubic to tetragonal phase transition for \ce{CsPbBr3}.
    Here, \emph{mean} denotes the average over all \gls{md} data while \emph{mode} refers to the point at which the probability is maximized.
    The solid lines correspond to the mean $\mu$ of the fits to \eqref{eq:double_gaussian}.
    The dotted line indicates the transition temperature between the orthorhombic and tetragonal phases.
    (c,d) Full distribution at \SI{302}{\kelvin}, indicated by the dashed line in a) and b), and solid lines indicate fits to \eqref{eq:double_gaussian}.
    (e--h) The same data as in (a--d) but for \ce{MAPbI3}, where the histograms are shown at \SI{309}{\kelvin}.
    }
    \label{fig:CsPbBr3-MAPI-transitions}
\end{figure*}

\textbf{Character of the transitions.}
Next, we address the character of the transitions in \ce{CsPbBr3}, which are analyzed via long (\SI{20}{\nano\second}), large-scale (compared to \emph{ab-initio} calculations; \num{61440} atoms) \gls{md} simulations in the $NPT$ ensemble in the vicinity of the transition temperature.
Energy, lattice parameters, and mode projections are recorded over \num{10} independent simulations for each temperature.
The energy referred to here is the potential energy after subtracting the zero Kelvin energy as well as the Dulong-Petit term $3\text{k}_\text{B}T/2$.

In the vicinity of the upper transition between the tetragonal and cubic phases, the histograms over the observed energy and lattice parameter values are multimodal (see \autoref{fig:CsPbBr3-MAPI-transitions}c,d for examples; see \autoref{sfig:CsPbBr3_distributions} in the \gls{si} for the complete data).
This behavior is due to the superposition of the distributions of the cubic and tetragonal phases, and the total histograms can be expressed as a sum of normal distributions (as shown in \autoref{fig:CsPbBr3-MAPI-transitions}c,d).
We emphasize that the two phases are not observed in the \gls{md} simulations \emph{simultaneously}; the system is rather switching between the two phases over time.

As a result of the multimodal character in the vicinity of the phase transition, the \emph{mean of the distribution} (circles in \autoref{fig:CsPbBr3-MAPI-transitions}a,b), which also corresponds to the time average of the respective quantity, changes gradually across the transition.\footnote{
    We note that the width of the coexistence region in temperature as it is observed here is affected by system size as well as the strength of the barostat.
    It should thus not be interpreted as an experimentally accessible quantity.
}

The \emph{mode of the distribution}, i.e., the position of the maximum (or most likely value), however, exhibits a discrete jump at the transition temperature (squares in \autoref{fig:CsPbBr3-MAPI-transitions}c,d), which is clear evidence for the first-order character of the transition.
The magnitude of the change in the potential energy yields a very small but finite latent heat of about \qty{0.6}{\milli\electronvolt\per\atom}.
This magnitude is consistent with a first-order transition with a (very) small free energy barrier separating the phases, allowing the system to jump between the two phases at the size and time scale of our \gls{md} simulations.
This barrier is also clearly visible in the mode coordinate of the M mode (\autoref{sfig:CsPbBr3_mode_coordinate}).

By contrast, for the lower (orthorhombic-tetragonal) transition, the distributions are unimodal and mean and mode coincide throughout the transition, which implies continuous character (\autoref{fig:CsPbBr3-MAPI-transitions}).

Our simulations thus show that the tetragonal-to-cubic transition is of first order while the orthorhombic-to-tetragonal is continuous, in agreement with experiments \cite{Hirotsu1974, Stoumpos2013}.
This is also consistent with our recent work on the phonon dynamics in this material, where the M-mode frequency in the cubic phase was observed to be non-zero at the tetragonal-to-cubic transition temperature \cite{FraRosEri23}.

\subsection{\texorpdfstring{\ce{MAPbI3}}{MAPbI3}}

\textbf{Behavior on heating/cooling.}
In the case of \ce{MAPbI3} we observe a transition during cooling at about \SI{300}{\kelvin} from the high-temperature cubic phase ($a^0a^0a^0$) to a tetragonal structure ($a^0a^0c^-$; \autoref{fig:CsPbBr3-MAPI-heating-cooling}c,d).
There are no further transitions as the system stays in the tetragonal phase down to \SI{0}{\kelvin}.

On heating, starting from the orthorhombic ground-state structure ($a^-a^-c^+$), a transition occurs at around \SI{150}{\kelvin} to a tetragonal (referring to $a=b \neq c$) structure\footnote{
    We refer to this phase as $a^-a^-c^0$ throughout this work to emphasize its relation to the other phases and its place in \autoref{fig:glazer-space-free-energy}a.
    We note, however, that in Glazer notation it could also be written as $a^0b^-b^-$.
} ($a^-a^-c^0$) followed by another transition at \SI{300}{\kelvin} to the cubic phase.

The cubic ($a^0a^0a^0$) and orthorhombic ($a^-a^-c^+$) phases as well as the tetragonal phase obtained on cooling ($a^0a^0c^-$) are observed experimentally, whereas the tetragonal phase observed on heating ($a^-a^-c^0$) is not.
This is a reflection of the stochastic nature of phase transitions.
Their outcome depends on the \emph{free} energy landscape at the transition temperature and is thus sensitive to both the free energy differences between possible structures and the barriers that separate them.
To understand the outcome of these simulations and the nature of the phase transitions (and pathways) in \ce{MAPbI3} it is therefore instructive to inspect the free energy landscape as a function of temperature.

\textbf{Free energy landscape in Glazer space.}
At \SI{350}{\kelvin} the free energy minimum is located at the center of the Glazer space, corresponding to the cubic phase ($a^0a^0a^0$, \autoref{fig:glazer-space-free-energy}c).
At \SI{300}{\kelvin} the minimum has shifted to the tetragonal structure with anti-phase tilting ($a^0a^0c^-$).
Similarly to the case of \ce{CsPbBr3} (\autoref{fig:glazer-space-free-energy}b), the free energy landscape close to the tetragonal-to-cubic transition is very wide and flat.

Interestingly, we also observe a local minimum in the free energy for a tetragonal $a^-a^-c^0$ structure at \SI{100}{\kelvin}.
Below \SI{100}{\kelvin} the free energy minimum shifts to the orthorhombic $a^-a^-c^+$ structure.
The experimental phase transition to the orthorhombic phase is around \SI{160}{\kelvin}, which is slightly higher than the transition temperature we find here.

\textbf{Character of the tetragonal-to-cubic transition.}
To gain further insight, we analyze the character of the tetragonal-to-cubic transition in \ce{MAPbI3} in the same manner as for \ce{CsPbBr3} (\autoref{fig:CsPbBr3-MAPI-transitions}).
We observe a discrete change in both energy (latent heat) and lattice parameters which indicates a first-order phase transitions.
The full distributions and corresponding fits are shown in \autoref{sfig:MAPI_distributions} of the \gls{si}.
The very small free energy barrier between these two phases makes the transition appear more continuous in \gls{md} simulations.
This is in agreement with the near tricritical character of this transition described by Whitfield \textit{et al.}, who based on experimental data reported coexistence between the cubic and tetragonal phase in the temperature range between 300 and \SI{330}{\kelvin} \cite{Whitfield2016}.
Here, we observe both phases in \gls{md} simulations in a narrow temperature interval between 305 and \SI{310}{\kelvin}.

\begin{figure}
    \centering
    \includegraphics{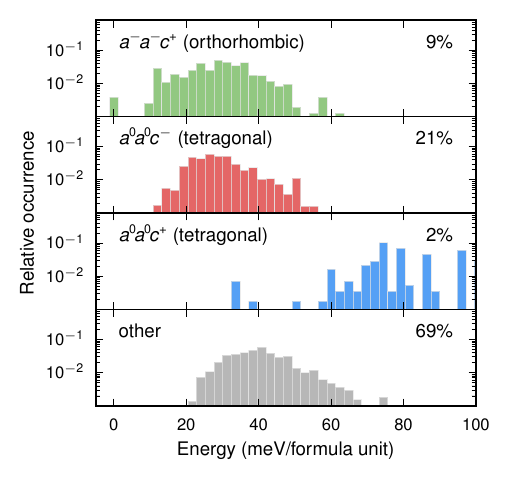}
    \caption{
        Energy distribution of fully relaxed \ce{MAPbI3} structures obtained by generating and relaxing \num{10000} possible tilted structures with randomized MA orientations in \numproduct{2x2x2} supercells of the primitive cubic cell.
    }
    \label{fig:MAPI_energies}
\end{figure}

\textbf{Molecular order and rotations.}
Replacing the Cs atom with the organic MA molecule gives rise to additional degrees of freedom and renders \ce{MAPI} significantly more complex than \ce{CsPbBr3}.
Crucially, the MA units have an orientation (C--N bond-vector), which is associated with ordering, in particular in the orthorhombic phase but also in the tetragonal phases \cite{quarti2014interplay, Weller2015, Lahnsteiner2016dynamics, Maheshwari2019, Sukmas2020, Cordero2021, Wu2023}.
These orientational degrees of freedom also make it much more challenging to sample the free energy landscape at low temperatures as they lead to energetic barriers between phases, which in turn allows metastable phases to remain dynamically stable down to \SI{0}{\kelvin}.
In order to understand the energetics of the \ce{MAPbI3} phases better we therefore conducted an extensive sampling of possible structures (\autoref{fig:MAPI_energies}).
\num{10000} different initial structures were constructed based on \numproduct{2x2x2} supercells of the primitive cubic unit cell, by applying a random set of tilt modes (either M or R) for each Cartesian direction with randomized amplitudes, and combined with randomized MA orientations.
Each structure was relaxed (atomic positions and cell shape) until the largest force fell below \SI{1e-4}{\electronvolt\per\angstrom}.
The relaxed structures were then classified into the Glazer structures by projection onto the M and R modes using a numerical tolerance of about \SI{0.5}{\degree}.

The first thing to note is that this approach correctly captures the $a^-a^-c^+$ structure as the ground state.
Furthermore, it is clear that the group of tetragonal $a^0a^0c^+$ structures is significantly higher in energy compared to $a^0a^0c^-$.
This is consistent with the fact that far fewer structures end up in the $a^0a^0c^-$ structure after relaxation.
Yet the majority of structures ends up at slightly higher energies (\SI{>20}{\milli\electronvolt\per\atom}) which illustrates the rather rough energy landscape of this material.

\begin{figure}
    \centering
    \includegraphics{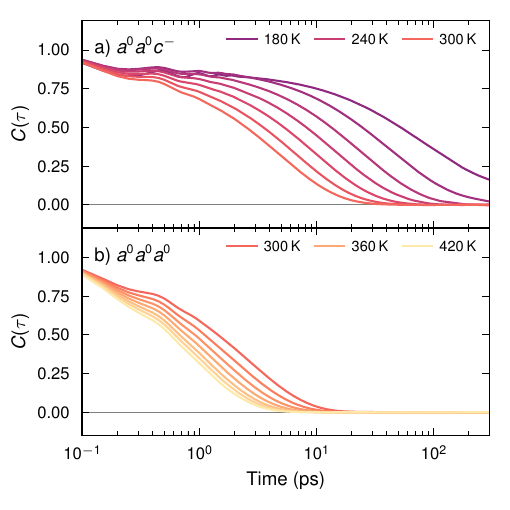}
    \includegraphics{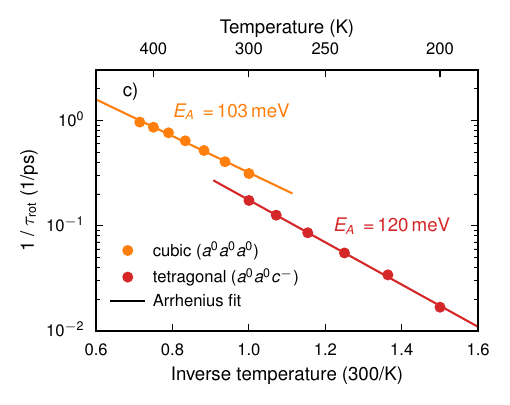}
    \caption{
        (a, b) \Acrlong{acf} $C(\tau)$ for the orientation of the \ce{MA} units in each phase.
        The spacing between the lines is \SI{20}{\kelvin}.
        (c) Rotation rate $1/\tau_\text{rot}$ obtained from (a--b) as a function of temperature.
        The solid lines correspond to Arrhenius fits $1/\tau_\text{rot}(T) \propto \textrm{e}^{-E_A/k_\text{B}T}$, where $E_A$ is the activation barrier of the rotation process.
    }
    \label{fig:MA_acfs}
\end{figure}

At finite temperature, the orientations of the \ce{MA} molecules fluctuate and can even flip at a significant rate even at modest temperatures \cite{Mattoni2015, Chen2015, Lahnsteiner2016dynamics, Fabini2017}.
This gives rise to a sizeable entropic contribution and a smoothing of the \emph{free} energy landscape with increasing temperature.
To analyze this aspect using the present approach, we ran \gls{md} simulations at several different temperatures starting from the lowest energy structure of each phase (\autoref{fig:MAPI_energies}) in supercells containing \num{96000} atoms.
Following an equilibration for \SI{1}{ns} in the $NVT$ ensemble (at the volume obtained previously from $NPT$ runs), then the C--N bond vector $\boldsymbol{r}_\mathrm{CN}(t)$ of each \ce{MA} unit was sampled in the $NVE$ ensemble (\autoref{sfig:MAPI_CN_distributions}).
We then analyzed the rotational dynamics of \ce{MA} in terms of the orientational \gls{acf}
\begin{align}
    C(\tau)
    = \frac{ \left < \boldsymbol{r}_\mathrm{CN}^i(t)
    \boldsymbol{r}_\mathrm{CN}^i(t+\tau) \right > } {
    \left < \boldsymbol{r}_\mathrm{CN}^i(t) \boldsymbol{r}_\mathrm{CN}^i(t) \right > },
\end{align}
where $\boldsymbol{r}_\mathrm{CN}^i(t)$ is the C--N bond vector at time $t$ for the $i$-th \ce{MA} unit.
For the orthorhombic structure ($a^-a^-c^+$), one observes very few \ce{MA} reorientation or rotation events on the time scales sampled here, and hence they are not included in the following analysis.
The \gls{acf} $C(\tau)$ exhibits oscillations on a shorter time scale (about \SI{100}{fs}) corresponding to atomic vibrations, and a slow decay on a longer time scale (about \SI{10}{ps} to \SI{100}{ps}) corresponding to \ce{MA} reorientation (\autoref{fig:MA_acfs}a,b).
The latter slow rotational behavior can be modeled with one exponential decay, $C(\tau) \propto \textrm{e}^{-\tau/\tau_\text{rot}}$, where $\tau_\text{rot}$ corresponds to a typical rotational time (see, e.g., \autoref{sfig:MAPI_CN_in_time}) \cite{Mattoni2015, Lahnsteiner2016dynamics}.
One can model the temperature dependence of the rotation time with an Arrhenius expression, $1/\tau_\text{rot}(T) \propto \textrm{e}^{-E_A/k_\text{B}T}$, which fits the data very well (\autoref{fig:MA_acfs}c).
This yields effective activation barriers $E_A$ for these rotational events of \SI{120}{\milli\electronvolt} and \SI{103}{\milli\electronvolt} for the tetragonal and cubic phases, respectively.
These activation barriers are in good agreement with experimental measurements \cite{Chen2015, Fabini2017}.

\begin{figure}
    \centering
    \includegraphics{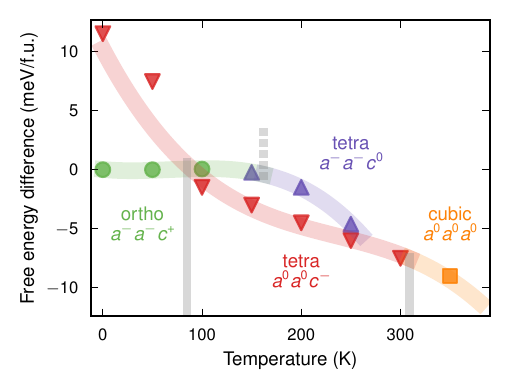}
    \caption{
        Phase diagram of \ce{MAPbI3} constructed based on \gls{md} simulations, free energy calculations, and energy minimization carried out in this work.
        Symbols represent local minima in the free energy landscape obtained from umbrella sampling (\autoref{fig:glazer-space-free-energy}).
        Lines are drawn as guide for the eyes.
        Free energy differences are given relative to the orthorhombic phase at each temperatures (suitably extrapolated at high temperatures).
        Grey horizontal lines indicate phase transition temperatures obtained from the free energy differences for the orthorhombic-to-tetragonal ($a^- a^- c^+$ to $a^0 a^0 c^-$) and from cooling \gls{md} simulations for the tetragonal-to-cubic ($a^0 a^0 c^-$ to $a^0 a^0 a^0$) transition.
        The dashed line indicates the transition from the orthorhombic $a^- a^- c^+$ to the tetragonal $a^- a^- c^0$ phase as observed in heating \gls{md} simulations (\autoref{fig:CsPbBr3-MAPI-heating-cooling}).
    }
    \label{fig:phase-diagram-MAPI}
\end{figure}

\textbf{Phase diagram and metastable phases.}
Finally, we can combine the information from the heating/cooling simulations (\autoref{fig:CsPbBr3-MAPI-heating-cooling} and \autoref{fig:CsPbBr3-MAPI-transitions}), the free energy landscape (\autoref{fig:glazer-space-free-energy}c), and the zero Kelvin energy distributions (\autoref{fig:MAPI_energies}) to construct a free energy diagram that shows the stability ranges of the different phases observed in this work (\autoref{fig:phase-diagram-MAPI}).

The tetragonal-to-cubic transition occurs around \SI{309}{\kelvin} according to the cooling simulations.
While it has first-order character the latent heat is very small, whence it can be readily sampled in \gls{md} simulations (\autoref{fig:CsPbBr3-MAPI-transitions}).
This is consistent with the free energy landscape in Glazer space, which shows the distinction between the tetragonal $a^0 a^0 c^-$ and the cubic $a^0 a^0 a^0$ phase to vanish slightly above \SI{300}{\kelvin} (
\autoref{fig:glazer-space-free-energy}c).

The transition from the orthorhombic $a^- a^- c^+$ to the tetragonal $a^0 a^0 c^-$ phase involves switching from in-phase to out-of-phase tilts relative to the $c$-axis and is unsurprisingly associated with a free energy barrier (\autoref{fig:glazer-space-free-energy}c).
This prevents us from sampling this transition directly with \gls{md} simulations.
Based on the free energy differences we can, however, locate this transition at approximately \SI{90}{\kelvin}.
As a result of this free energy barrier, the orthorhombic $a^- a^- c^+$ remains metastable up to about 150 to \SI{150}{\kelvin} (\autoref{fig:CsPbBr3-MAPI-heating-cooling}d).
Above this temperature it becomes unstable with respect to the alternative tetragonal $a^- a^- c^0$ phase, which itself is metastable with respect to the tetragonal $a^0 a^0 c^-$ phase that is also the one observed experimentally.
The metastable $a^- a^- c^0$ phase (alternatively written as $a^0b^-b^-$) is identifiable as a \emph{local} free energy minimum up to about \SI{250}{\kelvin} beyond which point the free energy differences become numerically too small to be resolvable.

While the present modeling does not yield the same transition temperature as experiments (caused by the approximations implicit to any current exchange-correlation functional \cite{FraWikErh2023}), we expect our results to be semi-quantitatively correct.
Hence, it is suggested that the metastability of the orthorhombic $a^- a^- c^+$ is a genuine feature of \ce{MAPbI3}, and the result of the orientational degrees of freedom associated with the molecular MA cation.
This might explain some of the uncertainty in the experimental analyses of this material.
Moreover it indicates the local structural variability that can be present as well as the complexity of the underlying dynamics.

\section{Conclusions}

For \ce{CsPbBr3} our simulations provide strong support for the cubic-to-tetragonal phase transition to have first order character, in agreement with the majority of experimental studies.
The latent heat is only on the order of a few \SI{0.1}{\milli\electronvolt\per\atom}, which allows one to observe the transition directly by \gls{md} simulations, even in rather large systems.
We find the orthorhombic-to-tetragonal transition in \ce{CsPbBr3} to be completely continuous, which overall gives rise to a rather simple free energy landscape and phase diagram.

As a result of the additional orientational degrees of freedom associated with the molecular MA cation, the situation is more complex for \ce{MAPbI3}.
While the cubic-to-tetragonal transition has first-order character with a similarly small latent heat as in the case of \ce{CsPbBr3}, the low temperature of the phase diagram is notably more intricate.
The orthorhombic-to-tetragonal transition has first-order character and is associated with a rather large free energy barrier that cannot be readily overcome on the length and time scales of \gls{md} simulations.
In our simulations the orthorhombic $a^- a^- c^+$ phase is found to be metastable up to about \SI{40}{\kelvin} above the orthorhombic-to-tetragonal phase transition.
We also identified a metastable tetragonal $a^- a^- c^0$ phase (alternatively written as $a^0b^-b^-$).

As noted before, the increased complexity of \ce{MAPbI3} compared to \ce{CsPbBr3} can be attributed to orientational degrees of freedom associated with the molecular cation.
This leads to an energy (and by extension free energy) landscape with many, usually rather shallow, local minima.
Most notably it causes the preferred tetragonal structure to switch from in-phase ($a^0 a^0 c^+$), which is preferred by many inorganic halide perovskites including \ce{CsPbBr3}, to out-of-phase ($a^0 a^0 c^-$) tilting.
This suggests that the trivial transition path from the tetragonal to the orthorhombic phase involves switching from out-of-phase to in-phase tilting, suggesting an intuitive understanding for the larger free energy barrier of this transition compared to transitions in inorganic halide perovskites such as \ce{CsPbBr3}.
The increased complexity of the (free) energy landscape moreover gives rise to metastable phases, which can in principle appear (possibly transiently) in experimental samples as well.
This might explain some of the uncertainty
in the experimental analyses of these materials \cite{druzbicki2016unexpected}.

Orientational degrees of freedom are also present in other hybrid (organic-inorganic) perovskites, most notably \ce{FAPbI3}, for which one can expect similar features.
It will therefore be particularly interesting to discriminate these effects in mixed systems, involving but not limited to \ce{MAPbI3}, \ce{FAPbI3}, and \ce{CsPbI3}.
We hope that the present work provides a strong basis for future explorations in these directions.

\section*{Acknowledgements}

This work was funded by the Swedish Research Council (grant numbers 2018-06482, 2019-03993, 2020-04935, 2021-05072), the Area of Advance Nano at Chalmers, and the Chalmers Initiative for Advancement of Neutron and Synchrotron Techniques.
J.~W. acknowledges the Swedish Strategic Research Foundation through a Future Research Leader programme (FFL21-0129).
The computations were enabled by resources provided by the National Academic Infrastructure for Supercomputing in Sweden (NAISS) and the Swedish National Infrastructure for Computing (SNIC) at C3SE, NSC, HPC2N, and PDC partially funded by the Swedish Research Council through grant agreements no. 2022-06725 and no. 2018-05973.

\section*{Supporting information}

The supporting information provides details pertaining to the \gls{dft} calculations (\autoref{snote:dft}) as well as the analysis of the character of the phase transitions (\autoref{snote:quantifying-the-transition-character}).
Moreover it contains figures illustrating the performance of the \gls{nep} model for \ce{MAPbI3} (\autoref{sfig:MAPI-parity}), the distribution of the mode coordinate (\autoref{sfig:CsPbBr3_mode_coordinate}), the analysis of the character of the phase transitions (\autoref{sfig:CsPbBr3_distributions} and \autoref{sfig:MAPI_distributions}), as well as the time dependence and distribution of the orientation of \ce{MA} units in \ce{MAPBI3} (\autoref{sfig:MAPI_CN_in_time} and \autoref{sfig:MAPI_CN_distributions}).
Finally, it contains a table comparing the energy differences from \gls{dft} calculations and \gls{nep} model for selected \ce{MAPbI3} structures (Table S1).

\section*{Data Availability}

The \gls{nep} model for \ce{MAPbI3} constructed in this study as well as a database with the underlying \gls{dft} calculations is openly available via Zenodo at \url{https://doi.org/10.5281/zenodo.8138960}.
The \gls{nep} model for \ce{CsPbBr3} is taken from Ref.~\citenum{FraRosEri23} and is available along with pertinent data via Zenodo at \url{https://doi.org/10.5281/zenodo.7313503}.

\end{document}